\documentclass[aps,prb,showpacs,twocolumn]{revtex4-1}

\usepackage{ulem}
\usepackage{graphicx}
\usepackage{CJK}
\usepackage{amssymb}
\usepackage{epsfig}
\usepackage{graphicx}
\usepackage{subfigure}
\usepackage{mathrsfs}
\usepackage{verbatim}
\usepackage{rotating}
\usepackage{longtable}
\usepackage{amsmath}
\usepackage{array,color}
\usepackage{dcolumn}
\usepackage{bm}

\begin{document}
\title{A response to commenter Ke Lan's comment on our paper published in Nature Communications (2023)14:5782 by J. Yan et al.}

\author{Ji Yan${ }^{1}$, Jiwei $\mathrm{Li}^{2,3}$, X. T. He${}^{2,3*}$, Lifeng Wang${ }^{2,3}$, Yaohua Chen${ }^2$, Feng Wang${ }^1$, Xiaoying Han${ }^2$, Kaiqiang Pan${ }^1$, Juxi Liang${ }^1$, Yulong Li${ }^1$, Zanyang Guan${ }^1$, Xiangming Liu${ }^1$, Xingsen Che${ }^1$, Zhongjing Chen${ }^1$, Xing Zhang${ }^1$, Yan Xu${ }^2$, Bin Li${}^2$, Minqing $\mathrm{He}^2$, Hongbo Cai${}^{2,3}$, Liang Hao${ }^2$, Zhanjun Liu${}^{2,3}$, Chunyang Zheng${ }^{2,3}$, M. Y. Yu${ }^5$, Shaoping Zhu${ }^{1,2}$}

\affiliation{${}^1$Laser Fusion Research Center, China Academy of Engineering Physics, 621900, Mianyang, P. R. China \\
${}^2$Institute of Applied Physics and Computational Mathematics, Beijing 100094, China \\
${}^3$Center for Applied Physics and Technology, Peking University, 100871, Beijing, P. R. China \\
${}^4$School of Physics, Peking University, 100871, Beijing, P. R.China \\
${}^5$College of Engineering Physics, Shenzhen Technology University, 518118, Shenzhen, P. R. China}
\collaboration{${}^*$xthe@iapcm.ac.cn}
\date{\today}
\maketitle

We recently read a comment article on arXiv\cite{Lan} criticizing our paper titled "Experimental Confirmation of Driving Pressure Boosting and Smoothing for Hybrid-Drive Inertial Fusion at the 100-kJ Laser Facility" by J. Yan et al., published in Nature Communications\cite{Yan2023NC}. The comment article, based on incorrect evidence and understanding, alleges that our paper is “not credible”. We have to write this response to clarify the commenter’s erroneous conclusions.

The main questions raised by Ke Lan in the comment article are:
\begin{enumerate}
    \item ``The authors only mention the pressure asymmetry of HD and ID, but not DD. This looks like the pressure asymmetry of DD can be completely ignored in the HD scheme. This is in odds ...";
    \item ``HD pressures do not inform us about the role of DD in the pressure amplification, which is not reported in this paper";
    \item ``The HD pressure should be superior to the sum of ID and DD pressure. Otherwise, why not use all the laser energy for DD-only or for ID only? Hence, to verify $\cdots$;
    \item The question is: ``how does the DD pressure compare to the ID and HD pressures."
    \item The commenter's conclusion is: ``Due to the lack of data on the direct-derive experiment under the same conditions, the logic of this paper is not close, so its conclusion is not credible" and ``The lack of experimental data on the direct-drive pressure leads us to the conclusion that the published experimental confirmation of driving pressure boosting and smoothing for hybrid- drive inertial fusion is not credible".
\end{enumerate}

Looking at the main questions above, the commenter’s key mistake is to think that the hybrid-drive (HD) pressure includes the direct-drive (DD) pressure, therefore alleged that “The lack of experimental data on the direct-drive pressure leads us to the conclusion that, ``the published experimental confirmation of driving pressure boosting and smoothing for hybrid-drive inertial fusion is not credible” and ``$\cdots$ the logic of this paper is not close, so its conclusion is not credible”. 

\textbf{This conclusion is not only untrue in content but also has logical problems.} In the content, unfortunately, the commenter did not earnestly read our paper published in Nature Communications \cite{Yan2023NC}, which clearly states that the HD pressure does not contain any DD pressure or any of its information thereof, and thus confused the concept of about what the HD pressure and the DD pressure are. On the other hand, logically, the commenter is also contradictory. Our experiments have confirmed boosting and smoothing of the HD pressure, showing that even if there is DD pressure in the HD pressure as said by the commenter, it has no affecting on the HD pressure, why do we need to do DD experiments separately to prove some properties of the DD pressure? Otherwise, ``$\cdots$ the logic of this paper is not close, so the HD pressure is not credible”, what kind of logic is that?

We believe that the commenter was likely confusing the completely different functions of the DD laser in the HD and DD schemes. In the HD scheme, the mass ablation rate for radiation is larger than that for electrons, and the indirect-drive (ID) laser and DD laser pulses act separately at different times. Consequently, the ablator surface expansion caused by the radiation ablation results in a much larger distance between the critical surface and the radiation ablation front. Thus, the supersonic electron heat wave converted by the DD laser near the critical surface propagates within this distance, becoming a compression wave before reaching the radiation ablation front. This compression wave is similar to a bulldozer, isothermally compresses the low-density plasma expanded from the radiation ablation front to a high-density. Thus, by increasing the plasma density while maintaining the high temperature, the radiation ablation pressure (the ID pressure) is converted into an HD pressure far higher than the ID pressure, as confirmed experimentally. This demonstrates that there is no DD pressure content or information involved.

In contrast, in the DD scheme, the mass ablation rate for electrons is much smaller than that for radiation, and the distance from the critical surface to the electron ablation front is much smaller than the distance from the critical surface to the radiation ablation front in the HD scheme. As a result, the supersonic electron heat wave converted by the DD laser near the critical surface does not have enough time to slow down to a compression wave with sonic speed as it does in the HD scheme. Instead, it directly acts on the ablator surface of the capsule, generating the electron ablation pressure (DD pressure) to implode the fuel in the capsule. The concepts of the HD pressure and the DD pressure will be further explained below.

\textbf{In the HD scheme}, there are two stages. In the first stage, only the ID laser works. The same as the ID scheme, the radiation ablation generates a radiation ablation pressure at the radiation ablation front on the surface of the ablator, which pre-compresses the capsule. At the same time, the high-temperature ablator surface expands in the coronal region to provide a sufficiently large distance $\Delta R_{I D}$ between the radiation ablation front and the critical surface. In the second stage, the ID laser continues to work, and the DD laser acts on the critical surface at the right time and is converted into a supersonic electron heat wave after absorption. It propagates and slows down within the large distance $\Delta R_{I D}$ in the ID local thermodynamic equilibrium (LTE, $T_e=T_i=T_r$ ) corona plasma, and slows down into a plasma compression wave before reaching the radiation ablation front. Meanwhile, the asymmetry of the supersonic electron heat wave propagating in the large distance $\Delta R_{I D}$ is well-smoothed. The compression wave is like a bulldozer, isothermally compressing the high-temperature, low-density plasma expanding from the radiation ablation front into a high-density plasma platform, thereby transforming the radiation ablation pressure into an HD pressure that is much larger than the radiation ablation pressure by increasing the plasma density and keeping the temperature unchanged. We can see from this that there is no DD pressure appearing there. \textbf{Although the DD laser in the HD scheme has the same name as in the DD scheme, its function is completely different from that in the DD scheme, so there will be no DD pressure in the HD pressure.}

\textbf{In the DD scheme}, since the mass ablation rate for electrons is much smaller than that for radiation, resulting in the distance between the electron ablation front and the critical surface much smaller than $\Delta R_{I D}$ in ID, the supersonic-electronic heat wave does not have time to slow down to the compression wave, and it acts on the ablator surface of the capsule to form an electron ablation pressure (DD pressure). What’s more, the small distance is also not sufficient to smooth the asymmetry of the supersonic-electronic heat wave.

For more information, the relevant HD content (see the fifth and seventh paragraph in our paper) has been extracted and attached in Appendix 1 in this response. A detailed explanation of DD pressure generating is attached in Appendix 2.

From the above explanation, we see clearly that in the HD scheme, there is no DD pressure involved in the process by increasing the density to achieve HD pressure boosting and smoothing in the large distance $\Delta R_{I D}$, and the HD pressure is not a sum of the ID pressure and the DD pressure. The purpose of the HD scheme is to generate a high enough and well-smoothed pressure. Therefore, the commenter's conclusion ``due to the lack of data on the direct-derive experiment under the same conditions, the logic of this paper is not close, so its conclusion is not credible.”, and ``The lack of experimental data on the direct-drive pressure leads us to the conclusion that the published experimental confirmation of driving pressure boosting and smoothing for hybrid-drive inertial fusion is not credible”, is groundless and completely wrong.

As for the DD pressure compared to the ID and HD pressures in which the three are independent of each other, we first explain that both the ID and DD pressures are ablation pressures, respectively, by radiation and electrons, and therefore they are only temperature dependent. The ID pressure (the radiation ablation pressure) at the radiation ablation front is $P_{I D} \propto T_r^{3.5}$, which easily is obtained from the balance of the radiation flux heating the surface of the ablator and the rate of work done, where $T_r$ is the radiation temperature. The DD pressure (the electron ablation pressure) at the electron ablation front is $P_{D D} \propto T_e \propto I_{D D}^{2 / 3}$ obtained from the balance of the electron flux heating the surface of the ablator and the rate of work done, where the electron temperature $T_e$ is proportional to $2 / 3$ power of the DD laser intensity $I_{D D}$. Different from the ID and DD pressures, the HD pressure is the plasma pressure that does not satisfy the balanced relationship, it is $P_{H D} \propto E_{D D}^{1 / 4} T_r$ depending on the DD laser energy $E_{D D}$ and radiation temperature $T_r$ and is nothing with the DD pressure $P_{D D} \propto I_{D D}^{{2}/{3}}$. For the $\mathrm{CH}$ ablator, ${P}_{I D}(\mathrm{M b a r}) \sim\left({4.5 - 5 )} ~T_r^{3.5}\right.$ with $T_r$ in the unit of $100 ~\mathrm{eV}, P_{D D}(\mathrm{M b a r}) \approx {9 0} I_{D D}^{2 / 3}$ with $I_{D D}$ in $\rm PW / \mathrm{cm}^2$, and $P_{H D}\mathrm{(M b a r)} \approx 180\left({E_{D D}}/{4}\right)^{{1}/{4}} T_r / 2$ with $E_{D D}$ in $\mathrm{kJ}$.

Taking $T_r=200~ \mathrm{eV}$, we have ${P}_{I D} \sim 45 ~\mathrm{Mbar}$ and taking $I_{D D}=1.8~ \mathrm{PW} /$ $\mathrm{cm}^2$, we have $P_{{D} {D}} \approx {1 3 3} ~ \mathrm{M b a r}$, they are nothing with DD laser energy $E_{D D}$. The HD pressure is nothing with the DD intensity $I_{D D}$. Although under the same $T_r=$ $200 ~\mathrm{eV}$, the experimental HD pressure ${P}_{{H D}}={180} ~ \mathrm{Mbar}$ for $\quad E_{D D}= 4 \mathrm{~kJ}$ is close to $\mathrm{P}_{\mathrm{DD}}=133~ \mathrm{Mbar}$ at $I_{DD}=1.8 ~\mathrm{PW} / \mathrm{cm}^2$, the $\mathrm{HD}$ pressure increases to ${P}_{{H}{D}}={850} ~ \mathrm{Mbar}$ when $E_{D D}=825 ~ \mathrm{kJ}$ while the DD pressure still is ${P}_{{D} {D}}= 133~ \rm Mbar$, for the ignition target. \textbf{This is why the HD pressure is superior to the ID and DD, not the summation of ID and DD pressures.}

\section{Appendix ~1}
The following is a detailed description of the hybrid-drive (HD) pressure, see the fifth and seventh paragraphs of our paper.

In the fifth paragraph of our article\cite{Yan2023NC}:

The hybrid-drive (HD) scheme, a coupling of ID and DD, in which the target consists of a spherical hohlraum rather than the cylindrical hohlraum and a layered DT fuel capsule with a CH ablator inside the spherical hohlraum, was proposed to provide an ideal HD pressure realizing the stable implosion and performing the hotspot ignition at the non-stagnation (before stagnation) time with the low convergence ratios and the suppression of hydrodynamic instabilities. In the whole HD implosion process, the ID laser with the pre-pulse and main pulse continuously enters the spherical hohlraum through laser-entrance-holes (LEHs) and is absorbed on the inner wall converting into thermal X-rays as schematically plotted in Fig. 1a. In the first phase, only the ID pre-pulse laser works. Due to the larger mass ablation rate ( $\dot{m} \propto T_r^3$ for the $\mathrm{CH}$ ablator) by radiation, the ablator surface produces a longscale-length ID corona plasma while the fuel in the capsule is pre-compressed by the asymmetric ID shocks driven by the nonuniform radiation ablation pressure. In the second phase, the ID main-pulse laser continuously offers the long-scalelength corona plasma and further enhances the pre-compression of the fuel. Meanwhile, the DD laser with intensity $I_L \sim 1-2 ~ \mathrm{PW} / \mathrm{cm}^2\left(1 ~\mathrm{PW}=10^{15} \mathrm{~W}\right)$ entering the spherical hohlraum along the opposite direction of the radius of the capsule is absorbed near the ID laser pre-offered critical surface and is converted to a supersonic-electronic-heat wave. We find that in the ID corona plasma, as long as the large enough distance $\Delta R_{I D}$ between the radiation ablation front and the critical surface is greater than close to a slowing-down length $d_s=\int_0^{\Delta t} v_e d t$, this supersonic-electronic-heat wave propagating in $\Delta R_{I D}$ can slow down to a sonic speed before reaching the radiation ablation front and a plasma compressive wave followed a precursor shock is formed, where $v_e$ is a supersonic-electronic-heat wave velocity and $\Delta t$ is the slowing down duration. This compressive wave with a high plasma pressure produces a novel effect under the stable support of the DD laser, similar to a ``bulldozer", to thermally compress the low ID corona plasma density $\rho_a$ near the radiation ablation front into sufficiently high HD plasma density $\rho_{H D}\left(\gg \rho_a\right)$ to form an HD density plateau between the compression wave front and the radiation ablation front, where the HD pressure of $P_{H D}=\Gamma \rho_{H D} T_r$ by increasing the plasma density is boosted much higher than the radiation ablation pressure $P_a=\Gamma \rho_a T_r$ with $\Gamma$ the ideal gas pressure constant. We find from numerical simulations that if $d_s$ matches with $\Delta R_{I D}$ well, in the density plateau there are fitted hydroscaling relations of the maximal HD pressure $\boldsymbol{P}_{\boldsymbol{H} \boldsymbol{D}} \propto \boldsymbol{E}_{\boldsymbol{D} \boldsymbol{D}}^{1 / 4} \boldsymbol{T}_{\boldsymbol{r}}$ and the maximal HD density $\rho_{H D} \propto E_{D D}^{1 / 4}$, where $E_{D D}$ is the DD laser energy.

On the other hand, during the supersonic electronic heat-conduction wave slowing down, its pressure nonuniformity $\frac{\delta P}{P}$ with the perturbation wavelength $\lambda_p=2 \pi R_{c r} / \ell$ caused by overlapping of DD laser beams near the critical surface of radius $R_{c r}$, is decayed and thermally smoothed very well in the form of $\frac{\delta P}{P} \approx$ $\left(\frac{2 \delta I_L}{3 I_L}\right)_{c r} \operatorname{Exp}\left(-\frac{2 \pi \beta \Delta R}{\lambda_{p}}\right)$ when $2 \pi \beta \Delta R>\lambda_p \quad\left(\right.$ or ${R_{c r}} / {(\beta \ell)} <\Delta R)$, where $\ell$ is the perturbation mode number and $\beta=1.5-2$ is a transverse thermal-ablation correction factor from 2D simulation, and this results in significant smoothing of the HD pressure.

In the seventh paragraph of our article\cite{Yan2023NC}: 

We have to explain why in the DD scheme there is no “bulldozer” effect because, in the HD scheme, before the DD laser arrives the thermal X-rays with the large mass ablation rate provided a long enough distance $\Delta R_{I D}$ between the radiation ablation front and the critical surface. While, in the DD scheme, there is no large enough slowing down distance prepared in advance, and the distance between the electron ablation front and the critical surface is too short due to the low mass ablation rate by electrons so that the supersonic-electron-heat wave directly hits on the capsule with no time to slow down.

\section{Appendix ~ 2}

In the DD scheme, the DD laser is absorbed on the critical surface and converted into an electron heat conduction wave, which propagates in the non-LTE electron plasma ( $T_e>T_i$ and radiation temperature $T_r \sim 0$ ), eventually directly heating the ``cold" high-density ablator surface, where there is an electron ablation surface (EAF). One side of the EAF is a rapidly expanding low-electron density corona region at a high electron temperature $T_e$. The reaction causes the other side to form a shock wave driven by a high-pressure $P_{D D}$ to implode the ``cold" high-density $\mathrm{CH}$ ablator and fuel. The driving implosion pressure on the EAF is $P_{D D}=$ $\Gamma_e \rho_e \mathrm{~T}_e$ with the electron density $\rho_e=2 \rho_c$ ( $\rho_c \approx 0.03 \mathrm{~g} / \mathrm{cc}$ is the critical density for the $\mathrm{CH}$ ablator at a $0.35 ~ \mu \mathrm{m}$ wavelength) obtained from the balance between the inward electron flux and the power of outward $\mathrm{PdV}$. Due to the electron temperature $\mathrm{T}_e \propto I_{D D}^{2 / 3}$ we get the DD pressure $P_{D D} \approx 90 I_{D D}^{2 / 3}$ Mbar, which is only related to the DD laser intensity $I_{D D}$ with $I_{D D}$ in units of $\mathrm{PW} / \mathrm{cm}^2$ rather than the HD pressure $P_{H D} \propto$ $E_{D D}^{1 / 4} T_r$ that depends on the DD laser energy $E_{D D}$ and the radiation temperature $T_r$.

\end{document}